\documentclass[conference]{IEEEtran}

\IEEEoverridecommandlockouts

\usepackage{cite}
\usepackage{amsmath,amssymb,amsfonts}
\usepackage{graphicx}
\usepackage{textcomp}
\usepackage{xcolor}
\def\BibTeX{{\rm B\kern-.05em{\sc i\kern-.025em b}\kern-.08em 		T\kern-.1667em\lower.7ex\hbox{E}\kern-.125emX}}

\usepackage{multirow}
\usepackage{tabularx,ragged2e,booktabs,caption}
\usepackage{epstopdf}
\usepackage{subcaption}
\usepackage{algorithm}
\usepackage{algpseudocode}
\usepackage{libertine}
\usepackage{amsthm}
\usepackage{mathtools}
\usepackage{balance}
\usepackage{pifont}
\usepackage{url}

\newtheorem{theorem}{Theorem}

\theoremstyle{remark}

\begin{document}

\title{Latency-aware and Survivable Mapping of VNFs in 5G Network Edge Cloud}

\author{\IEEEauthorblockN{Prabhu Kaliyammal Thiruvasagam, Abhishek~Chakraborty, and~C. Siva Ram Murthy}
	\IEEEauthorblockA{Indian Institute of Technology Madras, Chennai~600036, India\\
		prabhut@cse.iitm.ac.in, abhishek2003slg@ieee.org, murthy@iitm.ac.in} 
	\vspace*{-0.65cm}
}

\maketitle
\begin{abstract}
Network Functions Virtualization~(NFV) and Multi-access Edge Computing~(MEC) play crucial roles in 5G networks for dynamically provisioning diverse communication services with heterogeneous service requirements. In particular, while NFV improves flexibility and scalability by softwarizing physical network functions as Virtual Network Functions~(VNFs), MEC enables to provide delay-sensitive/time-critical services by moving computing facilities to the network edge. However, these new paradigms introduce challenges in terms of latency, availability, and resource allocation. In this paper, we first explore MEC cloud facility location selection and then latency-aware placement of VNFs in different selected locations of NFV enabled MEC cloud facilities in order to meet the ultra-low latency requirements of different applications (e.g., Tactile Internet, virtual reality, and mission-critical applications). Furthermore, we also aim to guarantee the survivability of VNFs and an edge server against failures in resource limited MEC cloud facility due to software bugs, configuration faults, etc. To this end, we formulate the problem of latency-aware and survivable mapping of VNFs in different MEC cloud facilities as an Integer Linear Programming~(ILP) to minimize the overall service provisioning cost, and show that the problem is NP-hard. Owing to the high computational complexity of solving the ILP, we propose a simulated annealing based heuristic algorithm to obtain near-optimal solution in polynomial time. With extensive simulations, we show the effectiveness of our proposed solution in a real-world network topology, which performs close to the optimal solution.
\end{abstract}


\section{Introduction}
\label{section_Introduction}

Network Functions Virtualization (NFV) and Multi-access Edge Computing (MEC)\footnote{Note that MEC is also known as Mobile Edge Computing. Also note that ``MEC'' and ``MEC cloud facility'' are synonymous and we use them interchangeably throughout this paper.} have emerged as promising key technology enablers for 5G networks and services. NFV replaces hardware middleboxes as Virtual Network Functions (VNF) that can be run on general purpose hardware, which increases flexibility and reduces capital and operational expenditures \cite{NFV_WP1_2012}. MEC enables network operators to support delay-sensitive services by moving cloud computing facilities from the core to the network edge \cite{MEC_WP1} \cite{MEC_WP2}.  The primary reason behind the introduction of MEC is to reduce the network latency and bandwidth consumption, and also to leverage the advantage of faster computing and decision-making at the edge of the access network. Since more data are generated at the edge of the network, processing data at the network edge can be efficient solution to accommodate more service requests with extreme service requirements. 

Although NFV and MEC based deployment enables to meet stringent and extreme delay requirements of future demands and to reduce capital and operational expenditures, still there are challenges that need to be addressed. In particular, availability, service continuity, and resource allocation are major concerns in MEC enabled 5G networks due to vulnerability in softwarization/cloudification of network functions, sharing of common resources to provide multiple services, and limited resources in MEC cloud facilities. VNFs are subject to failures due to software bugs, configuration faults, etc \cite{PKT_2020}. Random and unexpected failure of a VNF may disrupt the service abruptly and lead to Service Level Agreement (SLA) violations \cite{PKT_2019}. Hence, ensuring survivability of network functions is of paramount importance in 5G use case scenarios to guarantee the service continuity and to improve the quality of experience, which is one of the major requirements of 5G systems ~\cite{nfv2015} \cite{NFV_REL_003}.

\begin{figure}[]
	\centering
	\includegraphics[scale=0.24]{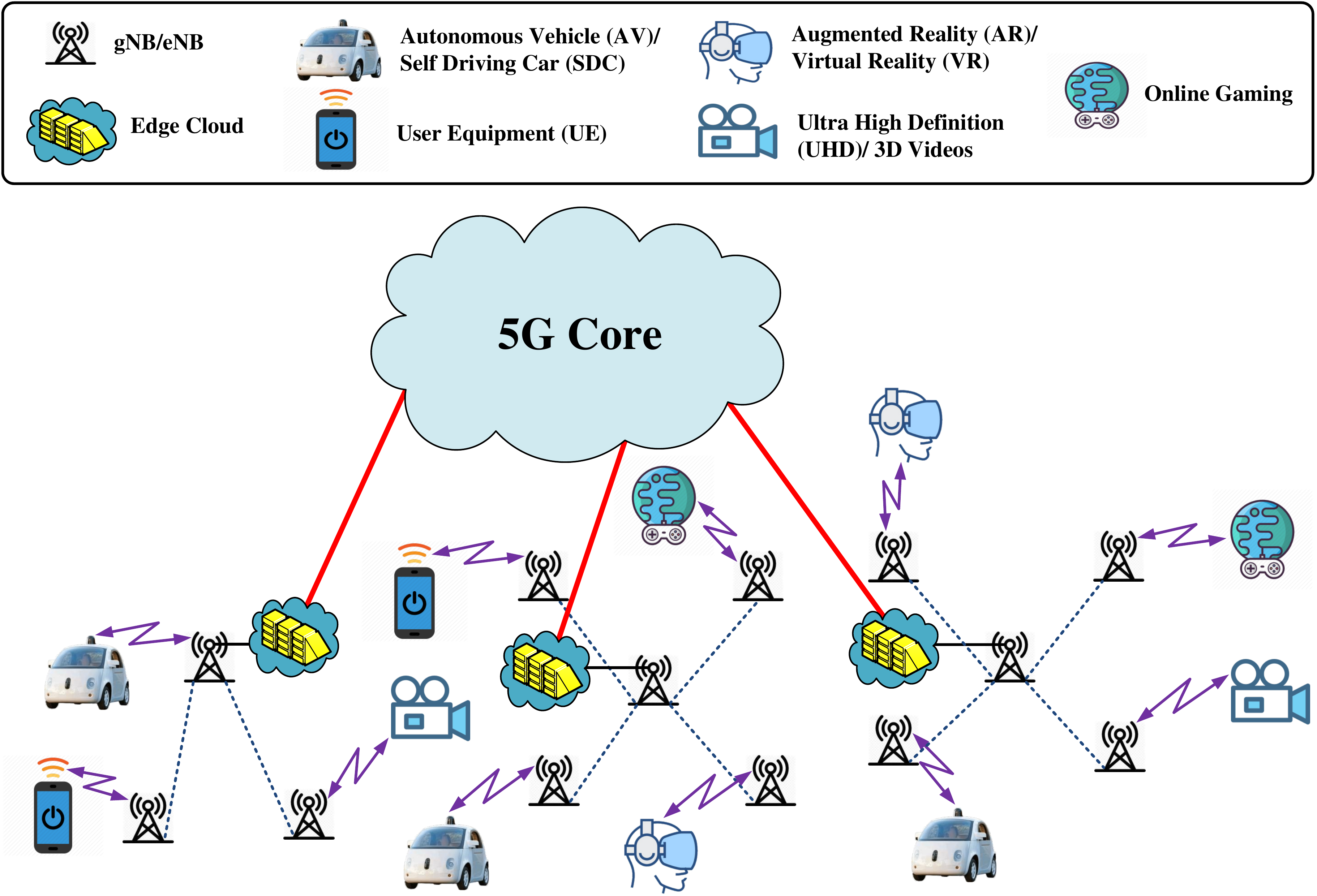}
	\caption{5G network architecture based on NFV and MEC, where a few MEC cloud facilities are placed at the selected influential locations.}
	\label{fig:MEC_Arch}	
\end{figure}

Fig.~\ref{fig:MEC_Arch} depicts a 5G network architecture that leverages the features of both NFV and MEC technologies. NFV enabled 5G core network is accessed through the Radio Access Networks~(RANs) in multihop fashion for providing services. However, NFV based softwarized infrastructure at the core cloud alone may not meet the communication delay requirement to process a service request from a distant source node, as it takes ample amount of time that may not be feasible in the context of latency-critical services such as autonomous driving and virtual reality-based services. Alternatively, edge cloud facilities can be established at the network edge for provisioning delay-sensitive services. As the service requests can be generated from any corner of a network, placement of a few MEC clouds near to the Base Stations (BSs) of RAN also demands attention \cite{MEC_WP2}. Therefore, MECs should be placed in such a way that the overall network can be covered in a manageable way and it also reduces the capital and operational costs ~\cite{li2018energy}. From Fig.~\ref{fig:MEC_Arch}, it can be observed that among the set of BS (eNB/gNB)  locations, only a few potential BSs~(in terms of influences based on the topological structure) are identified as the MEC cloud establishment locations. We note that the identified MEC cloud locations are very close to other remaining BSs in the network and, thus, reduce the response time for accessing a time-critical service. Hence, these MECs can be used to provide services in a timely manner. Furthermore, along with the proper selection of MECs, the required VNFs should also be optimally placed/instantiated on the MECs to provide the requested services efficiently. In this paper, we explore the problem of latency-aware and survivable mapping of VNFs in different selected locations of MEC cloud to provide delay-sensitive services in reliable manner.

The significant contributions of this paper are as follows:
\begin{itemize}
	
	\item We first propose an algorithm to select and establish a few MEC cloud facilities in the potential locations of BSs to cover the entire region and meet the delay requirements. 
	
	\item Then, we explore latency-aware placement of VNFs onto the servers of a few established MEC cloud facilities. In addition, to ensure survivability, we jointly place primary and backup VNFs in different edge servers, which enhances the availability of communication services and provides protection from random and unexpected failures.
	
	\item We formulate the problem as an ILP for provisioning services with minimal cost, and show that the problem is an NP-hard problem. 
	 
	\item We use CPLEX solver to find the optimal solution for the problem, and propose a simulated annealing based heuristic algorithm to provide near-optimal solution in polynomial time for large input instances. 
	
	\item With extensive simulations, we show the effectiveness of our proposed solution in a real-world network topology.

\end{itemize}

The remaining part of the paper is structured as follows: We review the related work in Section~\ref{section_RelatedWork}. We describe the system model and problem definition in Section~\ref{section_SystemModel}. We present MEC selection algorithm in Section~\ref{section_MECselection}. We formulate the ILP and present our proposed algorithm for latency-aware and survivable mapping of VNFs onto MEC edge servers in Section~\ref{section_ReliablePlacement}. We evaluate the performance of our proposed algorithms in Section~\ref{section_PerformanceEvaluation}. Finally, we conclude the paper with some directions for future work in Section~\ref{section_Conclusion}. 

\section{Related Work}
\label{section_RelatedWork}

In \cite{PKT_2020} and \cite{PKT_2021}, latency-aware and reliable VNF chain placement problem is considered, but MEC scenario is not taken into account. In \cite{Wang_2018}, an efficient VNF chain placement problem is considered in an MEC-NFV environment with the goal of maximizing the resource utilization. In \cite{Kiran_2020}, VNF placement and resource allocation problem is considered in NFV/SDN-enabled MEC networks with the goal of minimizing the overall placement and resource cost. In \cite{TS_2019}, VNF placement problem is considered at the network edges with the goal of minimizing end-to-end latency, and neural-network based model is used to proactively predict the number of VNFs required to process the network traffic. In \cite{RB_2019}, joint user association and VNF placement problem is considered for providing latency sensitive applications using MEC in 5G networks with the goal of minimizing the service provisioning cost. In \cite{Ben_2016}, QoS-aware VNF placement problem is considered in edge-central cloud architecture with the goal of efficiently allocating resources for provisioning services.
 
In \cite{Yala_2018}, latency-aware and availability driven VNF placement problem is considered in MEC-NFV environment with the goal of minimizing the cost. The work deals with availability of resources in MEC or core cloud and the latency associated with it. In \cite{Martini_2015}, latency-aware VNF composition problem is considered in 5G edge network with the goal of minimizing the overall latency. In \cite{Cziva_2018}, dynamic latency optimal VNF placement problem is considered at the network edge with the goal of minimizing the end-to-end latency.  In \cite{Emu_2020}, latency-aware VNF deployment problem is considered at edge for IoT services with the goal of minimizing the end-to-end latency. In \cite{Harris_2018}, latency-aware VNF placement and assignment problem is considered in MEC with the goal of maximizing the number of admitted service requests. In \cite{Chantre_2020} and \cite{Zhao}, resilient placement of VNFs in MEC is considered with the goal of minimizing the overall service provisioning cost. 

In the literature, MEC cloud network location is assumed to be given and the research is focused only on either latency-aware or resilient VNF placement in MEC. In this work, we first consider selection of MEC cloud facility location and then focus on both latency-aware and survivability aspects together in the placement of VNFs onto the selected/established MEC cloud facility with minimum cost. 

\section{System Model and Problem Definition}
\label{section_SystemModel}
We model the physical network as an undirected graph~$G =~(N,E)$, where $N$ denotes the set of BSs in the region and $E$ denotes the set of physical links that interconnect the BSs. The BSs can be interconnected by SDN based backhaul network. A small subset of BSs is chosen to establish MEC cloud facilities. We use the notation $L$ to denote the set of locations where MEC cloud network facilities are being established, where $L \subset N$. At each MEC cloud network facility, a set of limited number of servers $S$ are used to place VNFs in order to provide service for user requests. We use the symbol $C_s$ to denote the available resource capacity (e.g., CPU, RAM, and storage space) of each server $s \in S$. We consider a set of VNFs, denoted by $V$, that process data traffic to provide services for user requests. Each VNF $v \in V$ requires a certain amount of resource to  process the packets. The amount of resource required by VNFs is denoted by $C_v$ and it should be less than the available resource capacity of MEC cloud network servers. 

We consider that VNFs are subject to failures due to software bugs, configuration faults, unexpected failures of network functions, and cyber attacks (e.g., denial of service). Abrupt failure of a VNF may disrupt communication services and results in customer dissatisfaction and revenue loss. In order to enhance the reliability of communication services, backups are assigned to VNFs such that they meet SLAs and improve the service continuity.      

Multiple users are connected to the network through near by base stations and their service requests come through these base stations. 
We assume that each user service request $r \in R$ is represented as ($v^r, n^r, t^r, d^r$), where $v^r \in V$ denotes service type VNF, $n^r \in N$ denotes which BS user connects to and requests for service, $t^r$ denotes the data rate demand, and $d^r$ denotes the maximum allowed delay/latency. 

\noindent\textbf{Problem Definition:} Given a physical network graph $G$~=~($N,E$) and a set of service requests $r \in R$ with ($v^r, n^r, t^r, d^r$), find a joint efficient mapping of primary and backup VNFs~(for ensuring survivability) in different edge servers at MEC cloud facility locations to minimize the overall provisioning cost while meeting the SLAs.

\begin{algorithm}[]
	\begin{center}
		\centering
		
		\footnotesize
		
		\caption{MEC cloud facility location selection and establishment}
		\label{algo_MEC_Selection}
		\begin{algorithmic}[1]
			
			\Statex \textbf{Input}: Graph $G = (N, E)$ and maximum allowed delay requirement $D_{max}$ to reach MEC cloud facility  
			\Statex \textbf{Output}: Number of established MEC cloud facilities and its locations 
			\For{$i = 1\to |N|$}~{\Comment{\textit{Estimate the CC value of all BSs in ~$G$}}}
			\For{$j = 1\to |N|$}
			\State CC[i] = $\frac{1}{\sum_{ j}\text{distance(i,~j)}}$~{\Comment{\textit{distance(i,~j) is the shortest path distance \hspace*{3.7cm} between nodes i and j}}}
			\EndFor
			\EndFor 
			\State $S$ = Sort the nodes~(i.e.,~$N\in G$) in the descending order of the CC values
			\State Current node locations $L = \{l_1,l_2,......,l_{|N|}\}$ based on the CC value of nodes~ 
			{\Comment{\textit{the set of locations for establishing MEC cloud facility}}}
			\State Delay = $\infty$, $i$=1
			\While{Delay $\le D_{max}$} 
			\State Select the location $l_i$~{\Comment{\textit{high CC node location is selected}}} 
			\State Establish MEC cloud facility at the location $l_i$
			\State Delay = maximum delay from BSs in the network to reach one of \hspace*{0.45cm}the established MEC cloud facilities
			\State $i$=$i$+1
			\EndWhile		
			\State Return the number of MEC cloud facilities established and their corresponding locations	
		\end{algorithmic}
	\end{center}
\end{algorithm}

\section{MEC Selection and Establishment}
\label{section_MECselection}
In this paper, to identify a few potential MEC cloud facilities/locations $L \subset N$ \cite{MEC_WP2}, we select a set of influential BS nodes on the basis of high Closeness Centrality~(CC)~\cite{Freeman1978}\cite{Manoj2018} values. The procedure to select high influential BSs from the network is given in Algorithm \ref{algo_MEC_Selection}. Note that a node with a high CC value can be reached, from any distant node in a network, with a few hops~(or by traversing less distance). Therefore, we choose high CC-valued BSs as potential MEC cloud network locations.   

Algorithm~\ref{algo_MEC_Selection} first identifies the CC value of each BS to identify the influential nodes in the network (lines 1 to 5). Influential nodes are selected based on the connectivity and closeness with respect to all other nodes in the network. Then, a few MEC cloud facilities are established on the locations of high CC-valued BSs to reach MEC cloud facility from nodes in the network within the maximum allowed delay ($D_{max}$) requirement (lines~6 to~14). In this work, $D_{max}$ is set to 2 ms.

\section{ILP Formulation and Proposed Solution}
\label{section_ReliablePlacement}
\subsection{ILP Formulation}
\label{subsection_ILP_Formulation}

\noindent The objective is to place the required VNFs onto the MEC servers in reliable manner such that the placement strategy minimizes the overall provisioning cost while meeting the SLAs of diverse service requests.  

\begin{enumerate}
	\item \textbf{Decision Variables:} We define the following decision variables to formulate our problem of survivable placement of VNFs in MEC. 
	\begin{itemize}
		\item $w_{l}$: Binary variable that equals 1 if an MEC cloud facility $l \in L$ is chosen for providing service, and 0 otherwise. 
		\item $x_{ls}$: Binary variable that equals 1 if a server $s \in S$ is activated in the MEC cloud facility $l \in L$ to deploy VNF, and 0 otherwise.
		\item $y_{lsv}$: Integer variable that equals $\mathbb{N}$ if number of instances of VNF $v \in V$ are deployed on a server $s \in S$ in the MEC cloud facility $l \in L$, and 0 otherwise. 
		\item $y_{lsv^b}$: Integer variable that equals $\mathbb{N}$ if number of backup instances of VNF $v^b \in V$ are deployed on a server $s \in S$ in the MEC cloud facility $l \in L$, and 0 otherwise.  
		\item $z_{lsv}^{nr}$: Binary variable that equals 1 if a request $r \in R$ through the base station $n \in N$ is served by the VNF $v \in V$ which is placed on the server $s \in S$ in the the MEC location $l \in L$, and 0 otherwise.
		\item $z_{lsv^b}^{nr}$: Binary variable that equals 1 if a request $r \in R$ through the base station $n \in N$ is served by the backup VNF $v^b \in V$ which is placed on the server $s \in S$ in the the MEC location $l \in L$, and 0 otherwise.
	\end{itemize}
	
	\item \textbf{Objective Function:} The objective is to minimize the cumulative costs of number of physical MEC servers activated, number of VNFs deployed, and amount of traffic being forwarded on each link for provisioning reliable and delay-sensitive communication services.  
	
	i) Activation Cost of Physical MEC Server: It includes design, procurement, deployment, and maintenance costs of MEC cloud, where multiple servers are activated to host VNFs to provide reliable communication services. It can be expressed as follows: 
	\begin{equation}
		SC = c_{sc} \sum_{l \in L} \sum_{s \in S} x_{ls},
	\end{equation}
	where $c_{sc}$ denotes activation cost of a single server in MEC cloud locations.
	
	ii) Deployment Cost of VNF Instance: It includes the deployment/license cost of both primary and backup VNFs hosted on the physical MEC servers, which can be expressed as follows:
	\begin{equation}
		VC = c_{vc} \sum_{l \in L} \sum_{s \in S} \sum_{v, v^b \in V} (y_{lsv} + y_{lsv^b}),
	\end{equation} 
	where $c_{vc}$ denotes deployment cost of a VNF on any physical MEC server.
	
	iii) Forwarding Cost of Service Traffic: It is the cost for forwarding service request traffic from the base station of user to the MEC cloud facility where the VNF is hosted on the server to provide service, which can be expressed as follows:
	\begin{equation}
		TC = c_{tc} \sum_{l \in L} \sum_{s \in S} \sum_{v, v^b \in V} \sum_{n \in N} \sum_{r \in R} (z_{lsv}^{nr} + z_{lsv^b}^{nr}) \times t^r,
	\end{equation}
	where $c_{tc}$ denotes traffic forwarding cost for the service request $r$ and it is calculated per Mbps and $t^r$ denotes the data rate requirement of the service request.

	The objective is to minimize the overall cost of the aforementioned costs, which can be expressed as follows:
	\begin{equation}
		P: ~min~(\gamma_1 \times SC + \gamma_2 \times VC + \gamma_3 \times TC ),
		\label{obj_fun}
	\end{equation}  
	where $\gamma_1$, $\gamma_2$, and $\gamma_3$ are weighing factors to give relative importance to the objective functions. 
	
	\item \textbf{Capacity Constraints:} The resource requirement of VNFs should be within the limit of resources available in the MEC servers, and the processing capacity requirement of service requests should be within the limit of available processing capacity of VNFs.  
	
	i) The total resource requirement of VNFs (both primary and backup) to be placed should not exceed the available resource capacity of the MEC server which hosts VNFs. It can be expressed as follows:
	\begin{equation}
		\sum_{v, v^b \in V} C_v \times (y_{lsv} + y_{lsv^b}) \le C_s \times x_{ls}, \forall l \in L, \forall s \in S,
	\end{equation}
	where $C_v$ denotes resource requirement of VNFs and $C_s$ denotes the available resource capacity of the MEC server.
	
	ii) The total processing capacity requirement of service requests should not exceed the available processing capacity of VNFs~(both primary and backup), which can be expressed as follows: 
	\begin{equation}
		\sum_{n \in N} \sum_{r \in R} t^r \times z_{lsv}^{nr} \le pc_v \times y_{lsv}, \forall l \in L, \forall s \in S, \forall v \in V 
	\end{equation}
	
	\begin{equation}	
	\sum_{n \in N} \sum_{r \in R} t^r \times z_{lsv^b}^{nr} \le pc_v \times y_{lsv^b}, \forall l \in L, \forall s \in S, \forall v^b \in V 
	\end{equation}
	where $pc_v$ denotes the processing capacity of VNFs. 
	
	\item \textbf{Delay Constraint:} The delay requirement of service requests should be less than or equal to the delay between the service requesting base stations and MEC server locations, which can be expressed as follows:
	\begin{equation}
		\sum_{l \in L} \sum_{s \in S} \sum_{v \in V} d_{ln} \times z_{lsv}^{nr} \le d^r, \forall n \in N, \forall r \in R 
	\end{equation} 
	\begin{equation}
	\sum_{l \in L} \sum_{s \in S} \sum_{v^b \in V} d_{ln} \times z_{lsv^b}^{nr} \le d^r, \forall n \in N, \forall r \in R 
	\end{equation}
	where $d_{ln}$ denotes the communication delay between the service request carrying base station $n \in N$ and the MEC server location $l \in L$. 
	
	\item \textbf{Placement Constraint:} Each service request from the user through base station is assigned to two instances of the same VNF type (primary and backup) to ensure survivability, which can be expressed as follows:
	\begin{equation}
	\sum_{l \in L} \sum_{s \in S} \sum_{v \in V} z_{lsv}^{nr} = 1, \forall n \in N, \forall r \in R 		
	\end{equation}
	\begin{equation}
	\sum_{l \in L} \sum_{s \in S} \sum_{v^b \in V} z_{lsv^b}^{nr} = 1, \forall n \in N, \forall r \in R
	\end{equation}
	
	\item \textbf{Anti-affinity VNF Mapping Constraint:} The primary and backup VNFs should be placed in different edge servers in order to handle failure of VNFs or an edge server, which can be expressed as follows:
	\begin{equation}
		 \sum_{v,v^b \in V} (z_{lsv}^{nr} + z_{lsv^b}^{nr}) \le 1, \forall n \in N, \forall r \in R, \forall s \in S, \forall l \in L
	\end{equation}

	\item \textbf{Other Constraints:} \\
	 
	i) The MEC cloud location is chosen if at least one MEC server is activated in that location to place VNFs, which can be expressed as follows:
	\begin{equation}
		w_l = 1 ~\text{if}~ \sum_{l \in L} \sum_{s \in S} x_{ls}> 0, \forall l \in L
	\end{equation}	
	
	ii) The MEC server is activated if at least one VNF (either primary or backup) is placed on it, which can be expressed as follows:
	\begin{equation}
		x_{ls} = 1 ~\text{if}~ \sum_{v, v^b \in V} (y_{lsv} + y_{lsv^b}) > 0, \forall l \in L, \forall s \in S
	\end{equation}
	
	iii) VNFs (both primary and backup) are deployed if at least one service request from the user through base station requires the VNF to provide a particular service, which can be expressed as follows:
	\begin{equation}
		y_{lsv} = 1 ~\text{if}~ \sum_{n \in N} \sum_{r \in R} z_{lsv}^{nr} > 0, \forall l \in L, \forall s \in S, \forall v \in V		
	\end{equation} 
	
	\begin{equation}
	y_{lsv^b} = 1 ~\text{if}~ \sum_{n \in N} \sum_{r \in R} z_{lsv^b}^{nr} > 0, \forall l \in L, \forall s \in S, \forall v^b \in V		
	\end{equation} 
	 
\end{enumerate}

\begin{theorem}
Latency-aware and survivable mapping of VNFs in MEC is an NP-hard problem.	
\end{theorem}
\begin{proof}
	Let A be the problem of latency-aware and survival mapping of VNFs in MEC and B be the Reliable Capacitated Facility Location (RCFL) problem. RCFL problem is an optimization problem and it is NP-hard \cite{R-vMF5}. In RCFL problem, it is considered that facilities fail with equal probability and the model assigns primary and backup facilities for the demand to enhance the reliability. RCFL problem is defined as follows: the problem is to select facilities from the given set of potential facility locations, where each facility has limited capacity and subject to failure, to provide services to the demands such that the model is robust against failures and minimizes the cost of establishing facilities (primary and backup) and of transportation of goods from the facilities to the demand points. To prove that the problem A is NP-hard, it is sufficient to show that an instance of the problem B can be reduced to an instance of the problem A in polynomial time, i.e., B $\le_P$ A \cite{CormenAlgo}.   
	
	We can transform an instance of the problem B into an instance of the problem A in the following way: i) consider each facility in the problem B as equivalent to an MEC cloud facility in the problem A, ii) set the capacity of the facility in the problem B to be equal to the capacity of the MEC cloud facility in the problem A, iii) set the cost of activating facility in the problem B is equivalent to the activation cost of servers and deployment cost of VNFs (primary and backup) at MEC cloud in the problem A, and iv) set the transportation cost in the problem B as the traffic forwarding cost in the problem A. The transformation operation can be done in polynomial time of the input size. Hence, the problem B is reducible to the problem A in polynomial time. If A is not NP-hard, then B is also not NP-hard (since B is reducible to A), which is a contradiction. Therefore, it can be concluded that the problem A is also an NP-hard problem.      
\end{proof}

\subsection{Proposed Heuristic Solution}
\label{subsection_Heuristic}

\noindent As latency-aware and survivable mapping of VNFs onto the selected MEC cloud facility locations is an NP-hard problem, we develop a Simulated Annealing (SA) \cite{Kirkpatrick} \cite{ZM_DF} based heuristic algorithm to obtain near-optimal solution in polynomial time for providing services to user requests. Algorithm~\ref{algo_BatchMin} gives the procedure for latency-aware and survivable mapping of VNFs onto the MEC cloud facilities, and it is based on the concept of SA. SA is a probabilistic method for finding the global minimum cost function and the process may consist of multiple local minima. The SA algorithm is similar to traditional local search algorithms, but SA allows upward moves occasionally with the hope to come out of local minima. Although upward moves lead to increase in cost, it will help to escape from local minima. 

SA mathematically mirrors the physical process whereby a solid is slowly cooled to a frozen state of minimum energy. The minimum energy state (or ground state configuration) in statistical mechanics corresponds to the minimum cost function in combinatorial optimization problems, where the cost function plays the role of energy \cite{Kirkpatrick}.

\begin{algorithm}[]	
	\begin{center}
		\centering		
		\footnotesize 
		\caption{Simulated annealing based latency-aware and survivable mapping of VNFs onto the MEC cloud facilities}
		\label{algo_BatchMin}
		\begin{algorithmic}[1]
			\Statex \textbf{Input}: $G = (N, E)$ and a set of service requests with information ~($v^r, n^r, t^r, d^r$) $\forall r \in R$ 
			\Statex \textbf{Output}: Latency-aware and survivable mapping of VNFs onto the MEC cloud servers with minimum cost 
			\State $T = T_0$
			\State currentSol = Generate a current solution randomly
			\State Evaluate the current solution using the objective function (Equation \ref{obj_fun}), i.e., $c_1$ = cost(currentSol) 
			\While{$T > T_{min}$}
			\For{$i = 1\to \text{maxIterations}$}
			\State nextSol = Generate a next solution
			\If{nextSol does not violate any SLAs/constraints}
			\State Evaluate the next solution generated using the objective \hspace*{1.25cm}function (Equation \ref{obj_fun}), i.e., $c_2$ = cost(nextSol)
			\If{$c_2 \leq c_1$}
			\State currentSol = nextSol, i.e., $c_1 = c_2$
			\Else
			\State $r$ = random$(0,1)$
			\State $p$ = $e^{(c_1 - c_2)/T}$
			\If{$r < p$}
			\State currentSol = nextSol, i.e., $c_1 = c_2$
			\EndIf 
			\EndIf 
			\EndIf 
			\EndFor 
			\State $T = \alpha \times T$
			\EndWhile
			\State Return $c_1$			 
		\end{algorithmic}
	\end{center}
\end{algorithm}

In Algorithm~\ref{algo_BatchMin}, first an initial temperature is set and current feasible solution is generated and evaluated using the objective function (Equation \ref{obj_fun}). The current solution is generated by sorting the requests based on the latency requirement in ascending order and placing the required VNFs (primary and backup) onto different MEC cloud servers. We follow first fit principle to reuse the activated edge servers and deployed VNFs. Then, different next solutions are explored for maximum number of iterations. At each iteration in the inner loop, the cost of a new next solution is computed using the same objective function (Equation \ref{obj_fun}). If the cost difference ($c_1 - c_2$) is less than or equal to zero, then the solution is accepted directly, and the configuration of the new solution is set as the current solution. In the case that cost difference ($c_1 - c_2$) is greater than zero, a new solution is accepted with a certain probability. First, a random number $r$ is generated that is uniformly distributed between (0,1). Then, $r$ is compared with the probability value $p$ that is a function of the temperature and the cost difference of current and new solutions. If $r < p$, then the configuration of the new solution is set as the current solution; otherwise the original configuration is retained. At the end of maximum of number of iterations, the temperature value is updated. The process continues till the temperature goes below the minimum threshold temperature. The proposed heuristic follows the annealing process of cooling the temperature in a controlled manner and allowing bad movement with a certain probability to come out of a local minima.

\section{Performance Evaluation}
\label{section_PerformanceEvaluation}
In this section, we evaluate the performance of our proposed solution for solving latency-aware and survivable VNF mapping problem in MEC cloud networks. 

\subsection{Simulation Setup}
For the evaluation purpose, we use germany50 real-world network topology from SNDlib \cite{SNDlib10} which is a library of test instances for telecommunications network. The germany50 network topology consists of 50 nodes that are interconnected by 88 links. We assume that each node is a base station node. The selection of MEC cloud facility locations is based on closeness centrality metric from complex network theory as explained in Algorithm \ref{algo_MEC_Selection}, hence the chosen MEC cloud network locations are closely situated with respect to all other base stations in the network. 

We use CPLEX solver (version 12.8) and Concert Technology in Java to solve the ILP formulation, latency-aware and survivable mapping of VNFs in MEC. The proposed heuristic is implemented using Matlab and we run the simulations multiple number of times and take the average for evaluation. In the SA based heuristic design in Algorithm \ref{algo_BatchMin}, we set the initial temperature ($T_0$) as 100, the minimum temperature ($T_{min}$) for termination as 0.1, the number of inner loop iterations ($maxIterations$) as 50, and the cooling rate ($\alpha$) as 0.9. We have observed that going beyond 50 iterations do not improve the quality of the solution significantly.  

\subsection{Performance Analysis of MEC Cloud Facility Location Selection}
\begin{figure}[]
	\centering
	\includegraphics[scale=0.7]{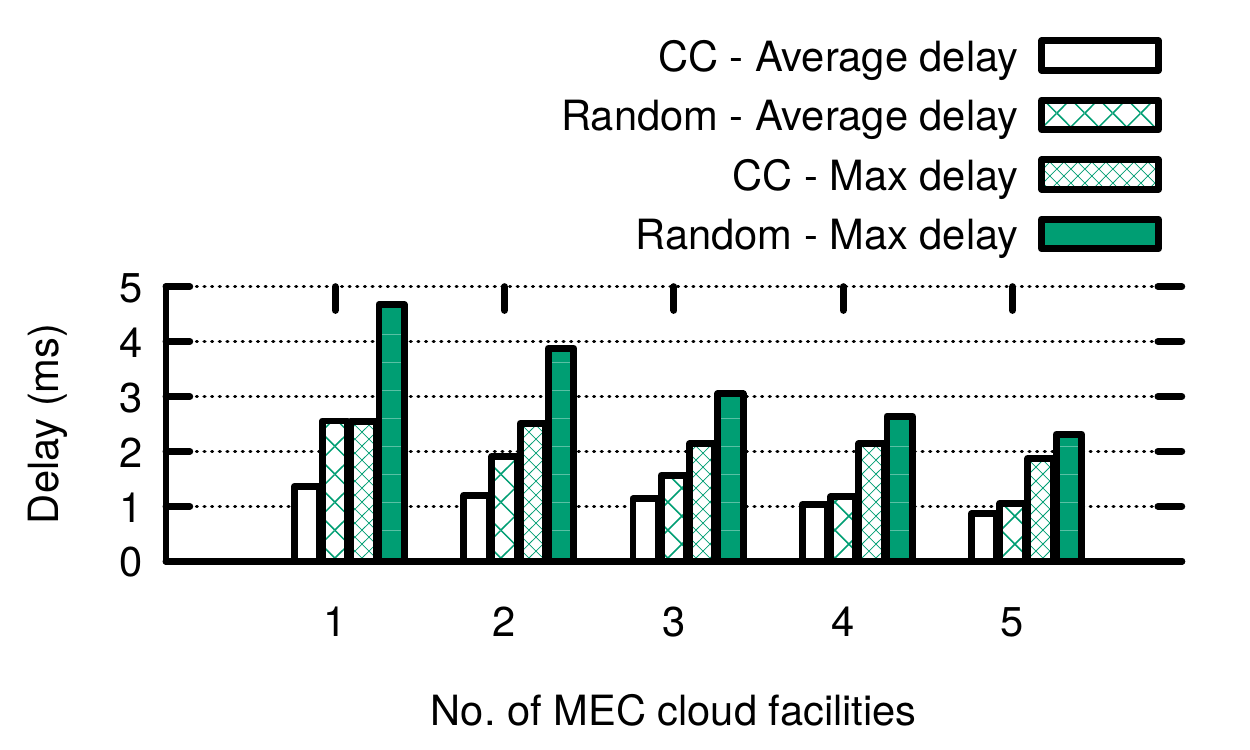}
	\caption{CC-based selection vs. Random selection.}
	\label{fig:CC_selection}
\end{figure}

As explained in Algorithm \ref{algo_MEC_Selection} earlier, MEC cloud facilities are chosen based on the centrality metric which depends on the topological structure of the network. We use CC to select the set of potential MEC cloud locations from germany50 network to provide latency-aware services. Fig.  \ref{fig:CC_selection} compares the performance of CC-based MEC selection with random selection. Average delay is the mean minimum delay to reach one of the MEC facilities from all the nodes in the network and max delay is the highest minimum delay from any node in the network to reach one of the MEC cloud facilities. As it can be seen that as we increase the number of MEC cloud facilities both average and max delays reduce. Since CC-based selection chooses high centrality nodes in the network, it takes less time for other nodes to reach the MEC facilities. Since random selection method  chooses the MEC facilities randomly and it may choose distant corner node as MEC cloud facility location, average and max delays are high compared to CC-based selection method. As it can be seen from Fig.  \ref{fig:CC_selection}, the maximum delay is within 2 ms when 5 MEC cloud facilities are established using CC-based selection method for providing services to users.    

\begin{table}[]
	\begin{center}
		\small
		\caption{Simulation parameters \cite{mec2018}}
		\label{tab:1}
		\begin{tabular}{|c|c|c|c|}
			\hline 
			\textbf{Service Types} & \textbf{Data Rate} & \textbf{Max Allowed Delay} \\
			\hline
			AR/VR & 200 Mbps & 2 ms \\
			V2X & 100 Mbps & 3 ms  \\
			e-health & 100 Mbps & 5 ms \\
			8K TV and Gaming & 200 Mbps & 10 ms \\
			\hline
		\end{tabular}
	\end{center}
\end{table}

\subsection{Performance Analysis of Latency-aware and Survivable Mapping of VNFs in MEC} 
We analyze the performance of our heuristic solution proposed in Section~\ref{section_ReliablePlacement} for solving latency-aware and survivable placement of VNFs in MEC with minimal cost. Table \ref{tab:1} shows the simulation parameters considered in this work, which are based on the requirements given in \cite{mec2018}. Four service types of VNFs are considered and each type has its corresponding data rate and maximum allowed delay requirements. We consider that each MEC cloud network has 10 MEC servers and each server has the resource capacity of 16 cores, and each VNF requires 4 cores and has the processing capacity of 1 Gbps \cite{power_aware}. Hence, 4 VNFs can be placed in a server and multiple services can share the same VNF. Each VNF service type has different data rate and maximum allowed delay requirement, and each user service request through the base station is randomly associated with one of the four service types with equal probability. The propagation delay between the base stations and MEC cloud network locations is computed based on the distance between them and considered that base stations are interconnected using optical fiber. We assume that processing and forwarding delay of the VNF is 50 $\mu s$ approximately \cite{Latency}. Our latency-aware VNF placement strategy satisfies the maximum allowed delay requirement by giving priorities to service requests of low delay requirements. For reliable service provisioning, service requests are associated with two different VNFs (active and backup) in different edge servers such that if the active VNF fails unexpectedly in random manner then the backup VNF takes charge to continue providing services without service interruption and disruption. From the above MEC cloud facility selection analysis, 5 MEC cloud facilities are enough to meet the required maximum delay requirement of 2 ms from the nodes in the network to reach one of the MEC cloud facilities.

\noindent Meeting the extreme requirements as well as effectively reusing the available resources to provide reliable services is a challenging task. The proposed algorithm aims to primarily meet the SLA requirements and reuse the shareable resources as efficiently as possible while ensuring survivability against failures.

We compare the performance of our proposed Simulated Annealing (SA) based heuristic solution against the following:
\begin{itemize}
	\item ILP: Formulated ILP problem is solved using CPLEX solver and it provides optimal solution.
	
	\item Greedy: This approach always places VNF in the nearest MEC cloud among all the possible MEC clouds which meet the SLA requirements. It incurs the minimal delay to provide services. 
	
	\item Baseline: This approach places VNF on the MEC cloud server that encounters first in the search space and satisfies the SLA requirements of the service request.
\end{itemize}

\begin{figure}[]
	\centering
	\includegraphics[scale=0.7]{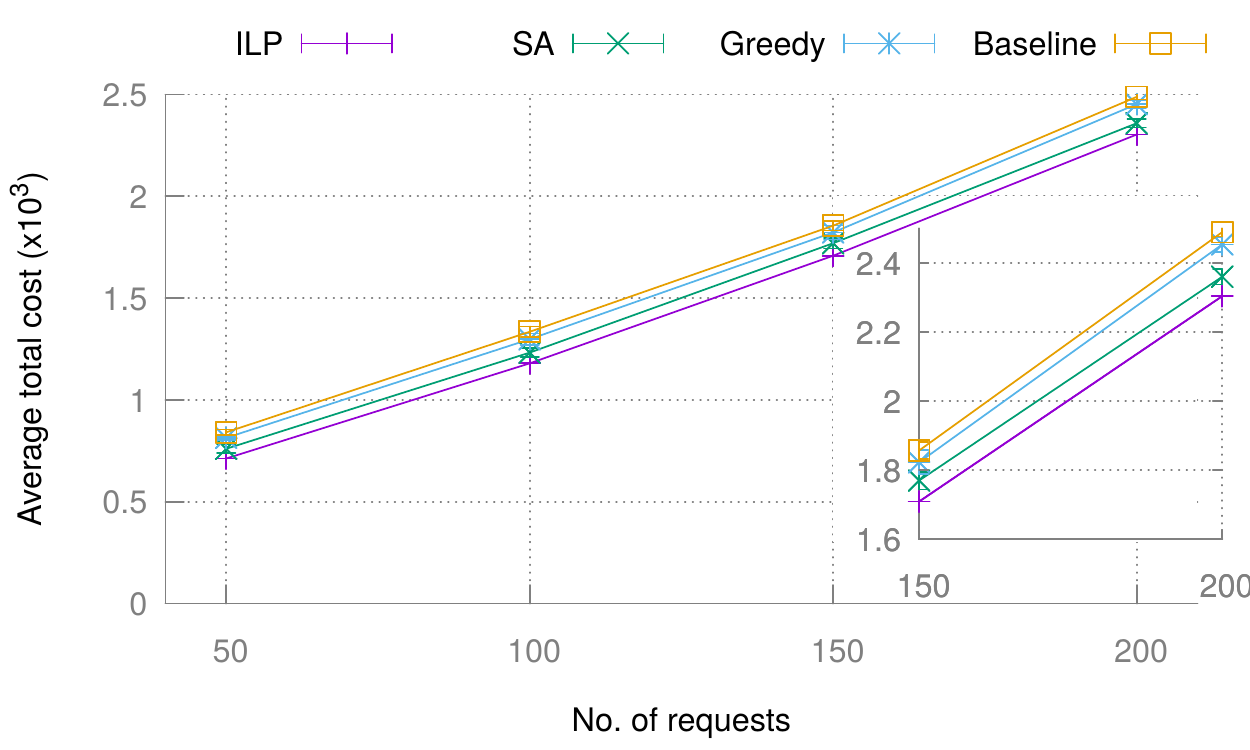}
	\caption{Comparison of average total cost for provisioning services.}
	\label{fig:comparison1_ILP}
\end{figure}

\begin{figure}[]
	\centering
	\includegraphics[scale=0.7]{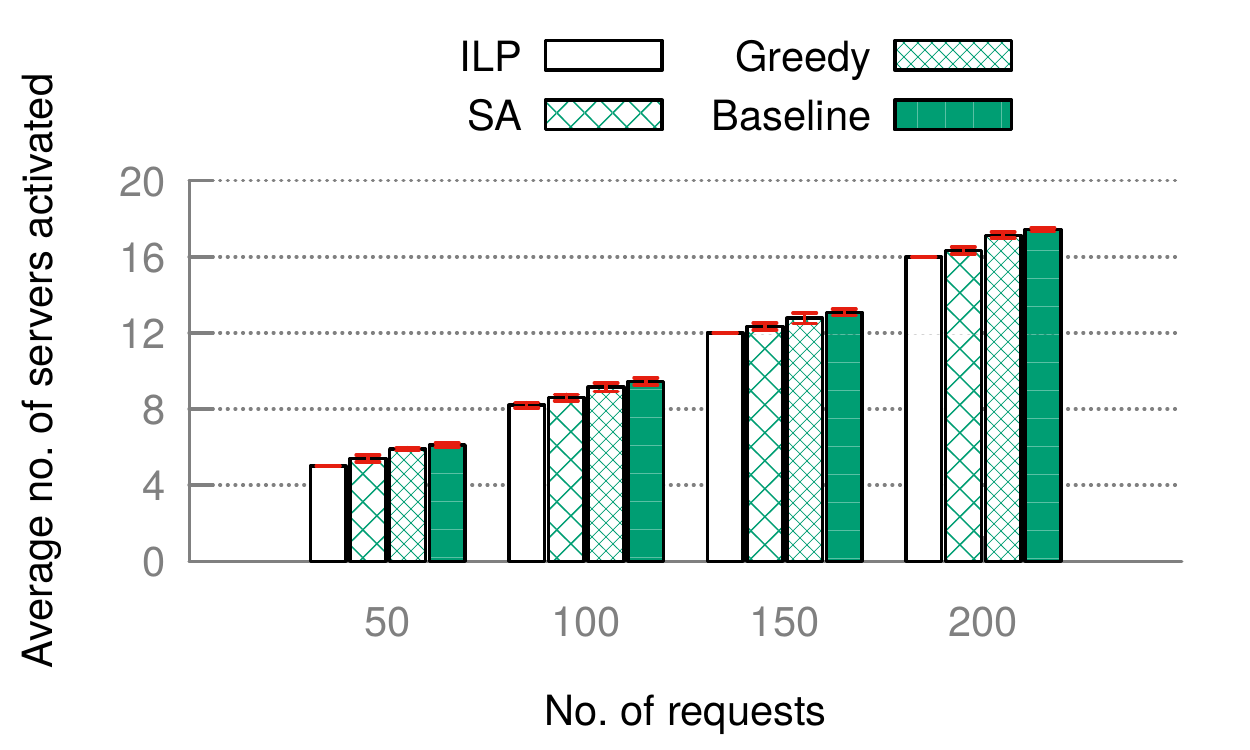}
	\caption{Comparison of average no. of MEC servers activated.}
	\label{fig:comparison2_MEC}
\end{figure}

\begin{figure}[]
	\centering
	\includegraphics[scale=0.7]{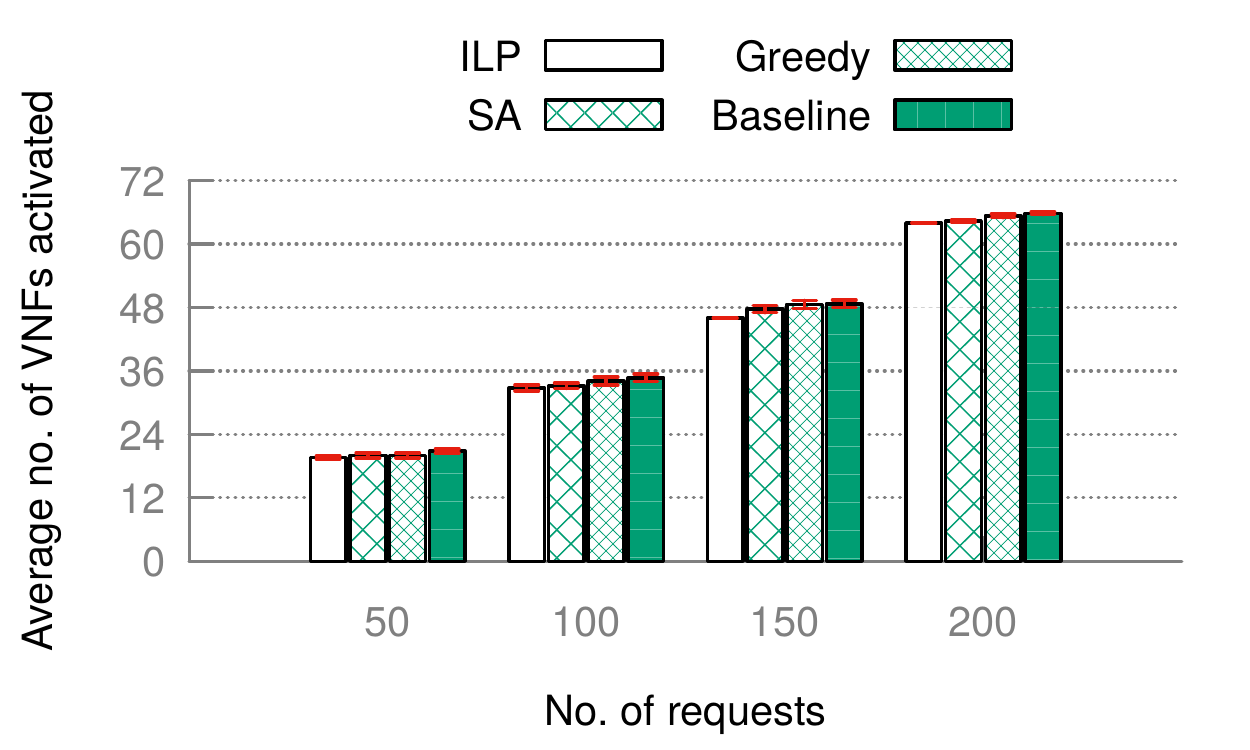}
	\caption{Comparison of average no. of VNFs activated.}
	\label{fig:comparison3_VNF}
\end{figure}

\begin{table}[t]
	\begin{center}
		\small
		\caption{Average running time comparison of different approaches (in seconds)}
		\label{tab:2}
		\begin{tabular}{|c|c|c|c|c|}
			\hline 
			\textbf{$\#$Service requests} & \textbf{50} & \textbf{100} & \textbf{150} & \textbf{200}  \\ \hline  
			ILP & 28.073  & 110.102 & 2298.805 & 21312.82  \\ \hline 
			SA  & 25.756 & 47.782 & 74.164 & 96.889  \\ \hline 
			Greedy  & 0.3549 & 0.4813 & 0.7592 & 0.8546  \\ \hline
			Baseline  & 0.4832 & 0.6288 & 0.8771 & 0.9981  \\ \hline			
		\end{tabular}
	\end{center}
	\vspace*{-0.15cm}
\end{table}

For the evaluation purpose, we consider that MEC cloud server activation cost is 100, VNF deployment cost is 10, and traffic forwarding cost is 1 per Mbps with respect to the delay between the service request carrying base station and the MEC cloud that provides service \cite{QoS}. Fig.  \ref{fig:comparison1_ILP} shows the comparison of total cost with respect to the number of requests being served. ILP takes the least cost for providing services. Although ILP provides the best solution, it is computationally expensive and impractical for large-scale input instances. Hence, we developed the Simulated Annealing (SA) based heuristic algorithm to provide near-optimal solution in polynomial time. We compare the performance of the proposed heuristic with the ILP (optimum solution).  As shown in Fig.  \ref{fig:comparison1_ILP}, our proposed SA based heuristic solution provides near-optimal solution (maximum optimality gap is 3.5\%) with lesser cost. Fig.  \ref{fig:comparison2_MEC} shows the comparison of average number of MEC servers activated to place the required VNFs, and it is clear that our proposed SA based heuristic solution performs close to the ILP and reduces the overall average cost as shown in Fig.  \ref{fig:comparison1_ILP}. Fig.  \ref{fig:comparison3_VNF} shows comparison of average number of VNFs activated. Although the average number of VNFs activated on different MEC servers are close for different approaches, the number of MEC servers activated to deploy the required VNFs for providing different classes of services to users differ clearly as shown in Fig.  \ref{fig:comparison2_MEC} and thus influences the overall cost for providing services.

\noindent We compare the average running time (in seconds) of ILP with different approaches for solving the latency-aware and survivable VNF placement problem in Table \ref{tab:2}. Solving ILP using CPLEX provides optimal solution in reasonable amount of time for small input instances. However, the running time to solve ILP increases exponentially as we scale the number of requests as shown in Table \ref{tab:2}. Owing to the high computational complexity of solving large instance of the ILP problem, we propose a SA based heuristic algorithm. The running time of heuristic algorithms are insignificant (in the order of seconds) compared to the running time of ILP. Since SA based heuristic algorithm explores different neighborhood solutions with respect to the temperature and the number of inner loop iterations, running time is in the order of seconds to provide near-optimal solution. Hence, from Table \ref{tab:2} it is clear that the proposed algorithm solves the problem in polynomial time. Since greedy and baseline approaches are executed once, they take less than a second to provide solution.

\vspace{-0.3cm}
\section{Conclusion}
\label{section_Conclusion}

In this work, we focused on latency-aware and survivable placement of VNFs in 5G network edge cloud. We first proposed an algorithm to select a few MEC cloud facility locations from the set of base stations to establish MEC cloud infrastructure and meet the user requirements of delay-sensitive services. Then, we explored both latency and survivability aspects together by leveraging the features of NFV and MEC cloud based technologies. We formulated the problem as an ILP to minimize the overall service provisioning cost (including both computing and communication resource cost). In order to overcome the high computational complexity of the ILP problem, we proposed a simulated annealing based heuristic algorithm which provides near-optimal and reliable solution to delay-sensitive heterogeneous service requests from users and industry verticals in polynomial time. We evaluated our proposed algorithm in terms of total provisioning cost and running time. With extensive simulations, we showed that our proposed solution performed close to the optimal solution (optimality gap is 3.5\%) in real-world network topology. 

In this work, we designed an offline algorithm to process the batch of service requests in order to place VNFs such that the SLA requirements are met. As a future work, we plan to design machine learning based online algorithm by considering the fact that the future service requests are not well known in advance. In addition, we would like to explore failure detection and rerouting mechanisms to analyze the actual delay incurred in the recovery process after failure of a network component in NFV/SDN-enabled 5G networks.  

\section*{Acknowledgement}
This research work was supported by the Department of Science and Technology (DST), New Delhi, India. 
 
\vspace{-0.11cm}
\bibliographystyle{IEEEtran}
\balance
\bibliography{MEC}

\begin{thebibliography}{10}
\providecommand{\url}[1]{#1}
\csname url@samestyle\endcsname
\providecommand{\newblock}{\relax}
\providecommand{\bibinfo}[2]{#2}
\providecommand{\BIBentrySTDinterwordspacing}{\spaceskip=0pt\relax}
\providecommand{\BIBentryALTinterwordstretchfactor}{4}
\providecommand{\BIBentryALTinterwordspacing}{\spaceskip=\fontdimen2\font plus
\BIBentryALTinterwordstretchfactor\fontdimen3\font minus
  \fontdimen4\font\relax}
\providecommand{\BIBforeignlanguage}[2]{{%
\expandafter\ifx\csname l@#1\endcsname\relax
\typeout{** WARNING: IEEEtran.bst: No hyphenation pattern has been}%
\typeout{** loaded for the language `#1'. Using the pattern for}%
\typeout{** the default language instead.}%
\else
\language=\csname l@#1\endcsname
\fi
#2}}
\providecommand{\BIBdecl}{\relax}
\BIBdecl

\bibitem{NFV_WP1_2012}
{ETSI NFV ISG}, ``{Network Functions Virtualization: An Introduction, Benefits,
  Enablers, Challenges and Call for Action},'' Oct. 2012.

\bibitem{MEC_WP1}
{ETSI ISG MEC}, ``Mobile edge computing: {A} key technology towards 5{G},''
  ETSI, Sep. 2015.

\bibitem{MEC_WP2}
------, ``{MEC in 5G networks},'' ETSI, June 2018.

\bibitem{PKT_2020}
P.~{Kaliyammal Thiruvasagam}, V.~J. {Kotagi}, and S.~R. {Murthy}, ``A
  reliability-aware, delay guaranteed, and resource efficient placement of
  service function chains in softwarized {5G} networks,'' \emph{IEEE
  Transactions on Cloud Computing}, pp. 1--1, 2020.

\bibitem{PKT_2019}
P.~K. Thiruvasagam, V.~J. Kotagi, and C.~S.~R. Murthy, ``The more the merrier:
  Enhancing reliability of {5G} communication services with guaranteed delay,''
  \emph{IEEE Networking Letters}, vol.~1, no.~2, pp. 52--55, June 2019.

\bibitem{nfv2015}
{ETSI GS NFV-REL 001 V1.1.1}, ``Network functions virtualization: Resiliency
  requirements,'' ETSI, Jan. 2015.

\bibitem{NFV_REL_003}
{ETSI GS NFV-REL 003 V1.1.2}, ``{Network Functions Virtualization; Reliability;
  Report on Models and Features for End-to-End Reliability},'' ETSI, July 2016.

\bibitem{li2018energy}
Y.~Li and S.~Wang, ``An energy-aware edge server placement algorithm in mobile
  edge computing,'' in \emph{Proc. IEEE International Conference on Edge
  Computing~(EDGE)}.\hskip 1em plus 0.5em minus 0.4em\relax IEEE, July 2018,
  pp. 66--73.

\bibitem{PKT_2021}
P.~K. {Thiruvasagam}, A.~{Chakraborty}, A.~{Mathew}, and C.~S.~R. {Murthy},
  ``Reliable placement of service function chains and virtual monitoring
  functions with minimal cost in softwarized {5G} networks,'' \emph{IEEE
  Transactions on Network and Service Management}, pp. 1--1, 2021.

\bibitem{Wang_2018}
M.~{Wang}, B.~{Cheng}, W.~{Feng}, and J.~{Chen}, ``{An Efficient Service
  Function Chain Placement Algorithm in a {MEC-NFV} Environment},'' in
  \emph{Proc. IEEE Global Communications Conference (GLOBECOM)}, 2019, pp.
  1--6.

\bibitem{Kiran_2020}
N.~{Kiran}, X.~{Liu}, S.~{Wang}, and C.~{Yin}, ``{{VNF} Placement and Resource
  Allocation in {SDN/NFV}-Enabled {MEC} Networks},'' in \emph{Proc. IEEE
  Wireless Communications and Networking Conference Workshops (WCNCW)}, 2020,
  pp. 1--6.

\bibitem{TS_2019}
T.~{Subramanya} and R.~{Riggio}, ``{Machine Learning-Driven Scaling and
  Placement of Virtual Network Functions at the Network Edges},'' in
  \emph{Proc. IEEE Conference on Network Softwarization (NetSoft)}, 2019, pp.
  414--422.

\bibitem{RB_2019}
R.~{Behravesh}, E.~{Coronado}, D.~{Harutyunyan}, and R.~{Riggio}, ``{Joint User
  Association and {VNF} Placement for Latency Sensitive Applications in 5{G}
  Networks},'' in \emph{Proc. IEEE International Conference on Cloud Networking
  (CloudNet)}, 2019, pp. 1--7.

\bibitem{Ben_2016}
F.~{Ben Jemaa}, G.~{Pujolle}, and M.~{Pariente}, ``{{QoS}-Aware {VNF} Placement
  Optimization in Edge-Central Carrier Cloud Architecture},'' in \emph{Proc.
  IEEE Global Communications Conference (GLOBECOM)}, 2016, pp. 1--7.

\bibitem{Yala_2018}
L.~{Yala}, P.~A. {Frangoudis}, and A.~{Ksentini}, ``{Latency and Availability
  Driven {VNF} Placement in a {MEC-NFV} Environment},'' in \emph{Proc. IEEE
  Global Communications Conference (GLOBECOM)}, 2018, pp. 1--7.

\bibitem{Martini_2015}
B.~{Martini}, F.~{Paganelli}, P.~{Cappanera}, S.~{Turchi}, and P.~{Castoldi},
  ``Latency-aware composition of virtual functions in 5{G},'' in \emph{Proc.
  IEEE Conference on Network Softwarization (NetSoft)}, 2015, pp. 1--6.

\bibitem{Cziva_2018}
R.~{Cziva}, C.~{Anagnostopoulos}, and D.~P. {Pezaros}, ``{Dynamic,
  Latency-Optimal {vNF} Placement at the Network Edge},'' in \emph{Proc. IEEE
  Conference on Computer Communications (INFOCOM)}, 2018, pp. 693--701.

\bibitem{Emu_2020}
M.~{Emu}, P.~{Yan}, and S.~{Choudhury}, ``{Latency Aware {VNF} Deployment at
  Edge Devices for {IoT} Services: An Artificial Neural Network Based
  Approach},'' in \emph{Proc. IEEE International Conference on Communications
  Workshops (ICC Workshops)}, 2020, pp. 1--6.

\bibitem{Harris_2018}
D.~{Harris}, J.~{Naor}, and D.~{Raz}, ``{Latency Aware Placement in
  Multi-access Edge Computing},'' in \emph{Proc. IEEE Conference on Network
  Softwarization and Workshops (NetSoft)}, 2018, pp. 132--140.

\bibitem{Chantre_2020}
H.~D. {Chantre} and N.~L. {Saldanha da Fonseca}, ``{The Location Problem for
  the Provisioning of Protected Slices in {NFV}-Based {MEC} Infrastructure},''
  \emph{IEEE Journal on Selected Areas in Communications}, vol.~38, no.~7, pp.
  1505--1514, 2020.

\bibitem{Zhao}
P.~{Zhao} and G.~{Dan}, ``Resilient placement of virtual process control
  functions in mobile edge clouds,'' in \emph{Proc. IFIP Networking Conference
  (IFIP Networking) and Workshops}, 2017, pp. 1--9.

\bibitem{Freeman1978}


\bibitem{Manoj2018}
B.~S. Manoj, A.~Chakraborty, and R.~Singh, \emph{Complex Networks: A Networking
  and Signal Processing Perspective}.\hskip 1em plus 0.5em minus 0.4em\relax
  Prentice Hall PTR, New Jersey, USA, Feb. 2018.

\bibitem{R-vMF5}
Z.-J.~M. Shen, R.~L. Zhan, and J.~Zhang, ``The reliable facility location
  problem: Formulations, heuristics, and approximation algorithms,''
  \emph{INFORMS Journal on Computing}, vol.~23, no.~3, pp. 470--482, 2011.

\bibitem{CormenAlgo}
T.~H. Cormen, C.~E. Leiserson, R.~L. Rivest, and C.~Stein, \emph{Introduction
  to Algorithms}.\hskip 1em plus 0.5em minus 0.4em\relax MIT Press, USA, July
  2009.

\bibitem{Kirkpatrick}
S.~Kirkpatrick, C.~D. Gelatt, and M.~P. Vecchi, ``Optimization by simulated
  annealing,'' \emph{SCIENCE}, vol. 220, no. 4598, pp. 671--680, 1983.

\bibitem{ZM_DF}
{Z. Michalewicz and David B. Fogel}, \emph{{How to Solve It: Modern
  Heuristics}}.\hskip 1em plus 0.5em minus 0.4em\relax Springer-Verlag,
  Germany, 2004.

\bibitem{SNDlib10}
S.~Orlowski, M.~Pi{\'o}ro, A.~Tomaszewski, and R.~Wess{\"a}ly, ``{SNDlib} 1.0
  -- {S}urvivable network design library,'' in \emph{Proc. International
  Network Optimization Conference~(INOC)}, Apr. 2007, pp. 1--15.

\bibitem{mec2018}
{ETSI White Paper}, ``{Cloud RAN and MEC: A Perfect Pairing},'' ETSI, Feb.
  2018.

\bibitem{power_aware}
A.~Varasteh, M.~De~Andrade, C.~M. Machuca, L.~Wosinska, and W.~Kellerer,
  ``Power-aware virtual network function placement and routing using an
  abstraction technique,'' in \emph{Proc. IEEE Global Communications
  Conference~(GLOBECOM)}, Dec. 2018, pp. 1--7.

\bibitem{Latency}
\BIBentryALTinterwordspacing
``{Network I/O Latency on VMware vSphere 5},'' 2020. [Online]. Available:
  \url{https://www.vmware.com/content/dam/digitalmarketing/vmware/en/pdf/techpaper/network-io-latency-perf-vsphere5-white-paper.pdf}
\BIBentrySTDinterwordspacing

\bibitem{QoS}
P.~{Vizarreta}, M.~{Condoluci}, C.~M. {Machuca}, T.~{Mahmoodi}, and
  W.~{Kellerer}, ``{QoS}-driven function placement reducing expenditures in
  {NFV} deployments,'' in \emph{Proc. IEEE International Conference on
  Communications (ICC)}, 2017, pp. 1--7.

\end{thebibliography}

\end{document}